%% file: main_preprint.tex
\documentclass[]{interact}

\usepackage{epstopdf}

\usepackage{siunitx}
\DeclareSIUnit{\belmilliwatt}{Bm}
\DeclareSIUnit{\dBm}{\deci\belmilliwatt}

\usepackage{subcaption}
\usepackage{graphicx}
\usepackage{booktabs}
\usepackage{multirow}
\usepackage{tikz}
\usepackage{tikz-qtree}
\usepackage{amsmath,amssymb,amsfonts}
\usepackage{mwe}
\usepackage{pifont}
\usepackage{url}
\usepackage{orcidlink}

\usepackage{natbib}
\usepackage{xcolor}

\usetikzlibrary{mindmap,shadows}
\usetikzlibrary{arrows.meta}
\usetikzlibrary{positioning}

\usetikzlibrary{arrows.meta, mindmap, shadows}
\usetikzlibrary{arrows}
\usetikzlibrary{backgrounds}

\pgfdeclarelayer{bg}    
\pgfsetlayers{bg,main} 

\bibpunct[, ]{(}{)}{;}{a}{}{,}
\renewcommand\bibfont{\fontsize{10}{12}\selectfont}

\theoremstyle{plain}

\theoremstyle{definition}

\theoremstyle{remark}

\newcommand{\p}[1]{{\text{\Pisymbol{psy}{#1}}}}

\usepackage{draftwatermark}
\SetWatermarkText{Preprint}
\SetWatermarkScale{5}

\begin{document}


\title{Simulation-based Evaluation of Indoor Positioning Systems in Connected Aircraft Cabins}

\author{
\name{Paul Schwarzbach\textsuperscript{a}\orcidlink{0000-0002-1091-782X}\thanks{CONTACT Paul Schwarzbach. Email: paul.schwarzbach@tu-dresden.de}, Jonas Ninnemann\textsuperscript{a}\orcidlink{0000-0001-7988-079X}, Hagen Ußler\textsuperscript{a}, Oliver Michler\textsuperscript{a}\orcidlink{0000-0002-8599-5304} and Michael Schultz\textsuperscript{b}\orcidlink{0000-0003-4056-8461}}
\affil{\textsuperscript{a}Chair of Transport Systems Information Technology, TUD Dresden University of Technology, Hettnerstr. 3, 01062 Dresden, DE; \textsuperscript{b}Chair of Air Traffic Concepts, University of the Bundeswehr Munich, Werner-Heisenberg-Weg 39, 85577 Neubiberg, DE}
}

\maketitle

\begin{abstract}
The manuscript discusses the increasing use of location-aware radio communication systems to support operational processes for the demanding aircraft cabin environment. In this context, the challenges for evaluation and integration of radio-based localization systems in the connected cabin are specifically addressed by proposing a hybrid deterministic and stochastic simulation approach, including both model-based ray-tracing and empirical residual simulation. The simulation approach is detailed in the manuscript and a methodology for applying and evaluating localization methods based on obtained geometric relations is conducted. This can also be used as a data generation and validation tool for data-driven localization methods, which further increase the localization accuracy and robustness. The derived location information can in return be used in order to perform operational prediction and optimization for efficient and sustainable passenger handling. A dataset derived from the introduced simulation platform is publicly available. 
\end{abstract}


\section{Introduction}

Efficient air traffic management and ground handling are crucial factors that drive the performance of airline operations. The integration of aircraft and passenger (PAX) trajectories presents a major challenge in future mobility management. The COVID-19 pandemic has significantly impacted the transportation network \cite{Schwarzbach_2020_Covid19}, highlighting the need for effective PAX handling processes at bottlenecks. PAX handling processes such as boarding and disembarkation within the confined aircraft cabin pose a challenge to air transportation. A key metric known as aircraft turnaround constitutes all ground-handling processes, including freight and baggage unloading and loading, refueling, cleaning, catering, and PAX boarding and disembarking \cite{schultz_future_2020}. PAX behavior largely influences the duration of boarding, and convenient and fast boarding is a priority for both PAX and airlines \cite{schultz_2018_implementation}. Queuing at boarding areas is a major concern for PAX. Airlines and airports aim to meet target times for ground operations provided by the responsible stakeholder, considering the current state of operations and experiences about the expected progress. Therefore, a high demand on both technological and operational solutions is present.

Location-aware radio-based communication networks provide a technical solution for PAX state monitoring and handling. While several general survey studies for these so-called indoor positioning systems (IPS), including \cite{Zafari_IPS_Survey_2019, MendozaSilva_meta_review_ips_2019, KimGeok_review_indoor_positioning_2020}, have been conducted, questions of technology selection, usable hardware, applicable positioning methods, as well as the holistic optimization of this emerging localization systems in the context of the connected cabin still remain unresolved. In addition, the accuracy and scalability of radio-enabled localization within the harsh propagation environment still poses a major research question. In terms of challenges for the development and integration of these systems within in the cabin environment, a variety of challenges arise. Among others this includes: The aircraft cabin is a restricted area so \textbf{accessibility} for integration and evaluation studies is limited. The \textbf{availability} of modern technologies, possible physical layer modifications or hardware components might not be available during field trials, which also causes limitations in the \textbf{adaptability} of the surveyed data, which is specific to the given scenario in which the data was surveyed.  

In order to evaluate the radio propagation channel within the aircraft cabin, a variety of empirical studies for different radio technologies have been conducted \cite{Chuang_cabin_radiowave_propagation_2007, Jacob_aircraft_human_body_uwb_2009, Chiu_cabin_uwb_human_presence_2010, schulte60GHzWLAN2011, Neuhold_experiments_uwb_aircraft_2016, Cogalan_cabin_radio_channel_measurements_2017}. The focus of these works are empirical findings concerning general characteristics of the propagation environment. 
In addition, localization capabilities of e.g. 5G \cite{cavdarDemonstrationIntegrated5G2018}, WiFi \cite{wangCooperativeLocalizationAssociation2021} or Ultra-Wideband (UWB) \cite{Karadeniz_aircraft_uwb_localization_2020, geyerPreciseOnboardAircraft2022} are discussed. However, these studies are limited to the technologies, constellation and hardware applied, so they allow no generalization.

To address the aforementioned challenges and shortcomings of conducted studies, an adaptive and technology-open simulation procedure is presented in this paper as simulation is typically applicable for early-stage evaluation. This is realized by the consolidation of both deterministic, model-based 3D ray-tracing \cite{Yun2015RayTracing} in order to derive signal characteristics, such as time-of-flight and visibility, and probabilistic error modeling in order to address additional hardware- or propagation-specific influences \cite{Schwarzbach_2021_statistical}. Unlike previous works, which focus on the general network coverage \cite{usslerDemoDeterministicRadio2022} or passive radio sensing \cite{Ninnemann_2022_multipath, Schwarzbach2022Enabling} within the cabin, this contribution focuses on the specialized demands and requirements of active localization using ray-tracing. However, these deterministic provide a strong correlation with the environment and geometry, as they are highly accurate but require complex modeling and computational efforts. These models are spatially consistent and can predict the channel's behavior with high accuracy. In addition, stochastic models are based on empirical channel measurements, and the distributions of signal parameters are determined for various reference environments. They can easily be implemented and do not require specific modeling of the environment. However, stochastic models may not be as accurate as deterministic models and may have some level of inconsistency in their spatial representation. To combine these advantages, the paper utilizes a hybrid simulation approach.

The manuscript presents the capabilities of radio-based simulation with respect to localization functionalities with applications for air transport management and especially in-cabin systems. The foundation for this is a ray-tracing respectively ray-launching based radio propagation simulation, which is commonly applied for network coverage analysis. In order to provide a feasible data basis for radio-based localization, the conventional simulation approach is modified. The work addresses the following main research objectives:

\begin{itemize}
    \item Presentation of a holistic radio propagation approach for active localization systems addressing the present issues of spatial consistency and the influence of stochastic errors by combining multiple simulation procedures;
    \item Merging of operational processes and simulations, e.g. boarding simulation with radio-based localization in a common framework;
    \item Highlighting the potentials of the simulation framework to evaluate also radio sensing applications and data-driven approaches.
\end{itemize}

The rest of the paper is structured as follows: Following this introduction, Section~\ref{sec:literature} provides a comprehensive overview on the operational aspects and potentials of location-awareness, as well as recent developments in radio-based localization and the corresponding radio channel modeling and simulation, with a specific emphasis on the connected aircraft cabin. Furthermore, Section~\ref{sec:methods} discusses the methodology of this work, including the proposed simulation portions and positioning capabilities. In Section~\ref{sec:results} the carried out simulation and respective results are discussed. The paper concludes with a conclusion and suggestions for future research directions.

\section{Literature Review}
\label{sec:literature}

The scope of the manuscript represents a cross-section of diverse scientific disciplines and their dedicated applications. These include operational planning and turnaround optimization within air traffic management, the realm of radio-based indoor positioning systems, as well as the modeling and simulation of these systems. Therefore, this section provides a brief literature review of the aforementioned research fields, including operational aviation research, radio-based localization as well radio chanel modeling, and highlights the resulting potential benefits of their integration.

\subsection{Operational Perspective}

Numerous studies have investigated the most efficient sequencing of PAX during boarding and deboarding processes \cite{2005_Ferrari_Robustness,van_landeghem_reducing_2002,tang2012aircraft,milne2018robust}. These studies typically analyze the effects of various operationally relevant input parameters, including seat load factor, group composition (such as families and couples), volume of carry-on baggage, and use of baggage compartments. The objective of optimizing boarding/deboarding processes is generally to minimize the amount of time required. The duration necessary for establishing the sequence or the stability of the solution is commonly disregarded (see \cite{schultz_2018_implementation} for variability in boarding time). 

Since the PAX cabin lacks sensors, optimization strategies cannot consider the current situation in the cabin. Therefore, a sequence often deemed advantageous might not be the optimal solution in a specific scenario. Implementing a cabin sensor system would facilitate the collection of movement data (such as walking speeds and position of bottlenecks~\cite{schultz_fast_2018}) and status data regarding cabin configuration (such as seat occupancy and baggage compartment usage). 

As currently, sparse data on cabin processes are available, they could be utilized to calibrate airline-specific models~\cite{schultz_fieldTrial_2018} and support dynamic optimization approaches~\cite{notomista2016fast,zeineddine_2017_dynamic}. These dynamic approaches can then update the sequence considering the current congestion situation. If congestion makes the rear rows to board problematic, priority should be given to front-row seat PAX. Similarly, data on the overhead storage compartments could be utilized. Operating at total capacity, these compartments indicate that nearby PAX may take longer to stow their own luggage, potentially obstructing the aisle for an extended duration. In addition, the data of the cabin conditions can be recorded, used to develop data-driven models, and finally predict the process flows, e.g., prediction of boarding times ~\cite{schultz2019machine}.

With the COVID-19 pandemic, new challenges have emerged for cabin operations that focus not only on boarding time but also on minimizing the risk of virus transmissions \cite{milne2021airplane,schultz_2021_covid_solve}. Given these new operational requirements, sensor-based location tracking of PAX within the cabin is an important requirement and no longer just a desired feature \cite{Schwarzbach_2020_Covid19}.

\subsection{Radio-based Localization}

The determination of an absolute location information can be realized using radio-based systems. In this context, radio devices that can be distinguished as stationary infrastructure units (anchors) and mobile tags to be located. In principle, range-based or range-free relations are derived from physical parameters of the systems, which can then be used to calculate the tag's location with knowledge of the anchor's location \cite{MendozaSilva_meta_review_ips_2019}.

As a technological basis, wireless communication systems have garnered significant research attention over the past two decades due to their cost-effectiveness, energy efficiency, retrofitting capabilities, and versatility \cite{Zafari_IPS_Survey_2019}. This has led to the rapid emergence of the Internet of Things (IoT) across various research and application domains, enabled by affordable smart interconnections of devices, individuals, and operations \cite{AlFuqaha_2015_IOT_Survey, lin_iot_survey_2017}. Researchers have explored using radio-based localization technologies to enhance intelligent and automated systems in various application areas, going beyond fundamental IoT functions like communication and sensor data acquisition\cite{Li_location_enabled_iot_survey_2021}. Similar to the utilization of Global Navigation Satellite Systems (GNSS) \cite{EgeaRoca_gnss_state_of_the_art_trends_2022}, the widespread adoption of smartphones \cite{Xiao_survey_indoor_positioning_2016, Davidson2017IPSSurvey} and connected wearables \cite{Ometov_survey_wearable_2021} has driven the comprehensive use of location-enabled communication technologies.

Terrestrial location-aware communication systems, historically confined to GNSS-denied scenarios, are commonly referred to as Indoor Positioning Systems (IPS) \cite{Farahsari2022}. An additional research focus involves the ongoing transformation of mobile communication systems, particularly with the advent of 5G New Radio technology\cite{delPeralRosado_survey_localization_cellular_radio_1g_5g_2018, chettri_iot_survey_5G_2020} and future  systems \cite{Kanhere_5G_beyond_positioning_2021, Chen_tutorial_terahertz_localization_6g_2022}. These developments have transitioned from conventional assistance of positioning using signals of opportunity to fully integrated communication systems \cite{Liu_2022_fundamental_limits_isac}. 

In this context, integrated communication systems encompass the ability to provide both active, device-based localization and passive, device-free sensing \cite{Shastri_review_mmwave_localization_2022}.

A plethora of survey papers on location-awareness and location capabilities in IPS and mobile cellular networks are available \cite{Gu_survey_ips_wpn_2009, Xiao_survey_indoor_positioning_2016, Tariq2017, Ferreira2017IPSSurvey, Laoudias_survey_radio_technologies_localization_tracking_navigation_2018, Zafari_IPS_Survey_2019, KimGeok_review_indoor_positioning_2020, Shastri_review_mmwave_localization_2022}. Analyzing these works reveals a lack of a unified taxonomy for defining location-aware communication systems. Instead, existing works are typically driven by either technology or application considerations. However, the abundance of research indicates that technologies, applications, and regulations are subject to continuous evolution. Concerning the establishment of a framework for spatial and geometric relations within a radio network, abstracted unified measurement types for data acquisition are necessary. These encompass measurement principles related to signal characteristics (e.g., time-based), physical attributes (e.g., frequency bands, bandwidths), and connection-related properties (e.g., network topology, directionality) of communication technologies.

In light of the aforementioned aircraft cabin environment, several empirical studies have been conducted to assess radio channel and propagation characteristics \cite{Chuang_cabin_radiowave_propagation_2007, Jacob_aircraft_human_body_uwb_2009, Chiu_cabin_uwb_human_presence_2010, schulte60GHzWLAN2011, Neuhold_experiments_uwb_aircraft_2016, Cogalan_cabin_radio_channel_measurements_2017}. These investigations also explored the localization capabilities of various radio technologies, including Ultra-Wideband (UWB) \cite{Karadeniz_aircraft_uwb_localization_2020, geyerPreciseOnboardAircraft2022}, WiFi \cite{wangCooperativeLocalizationAssociation2021}, and 5G \cite{cavdarDemonstrationIntegrated5G2018}. These studies primarily aim to provide empirical insights into radio propagation in the challenging aircraft cabin environment. However, their focus on assessing localization capabilities have been somewhat limited. Consequently, there remains a need for integrated communication systems that facilitate precise location determination through radio-based positioning within such environments.

\begin{figure}[htb!]
    \centering
    \includegraphics[width=1\linewidth]{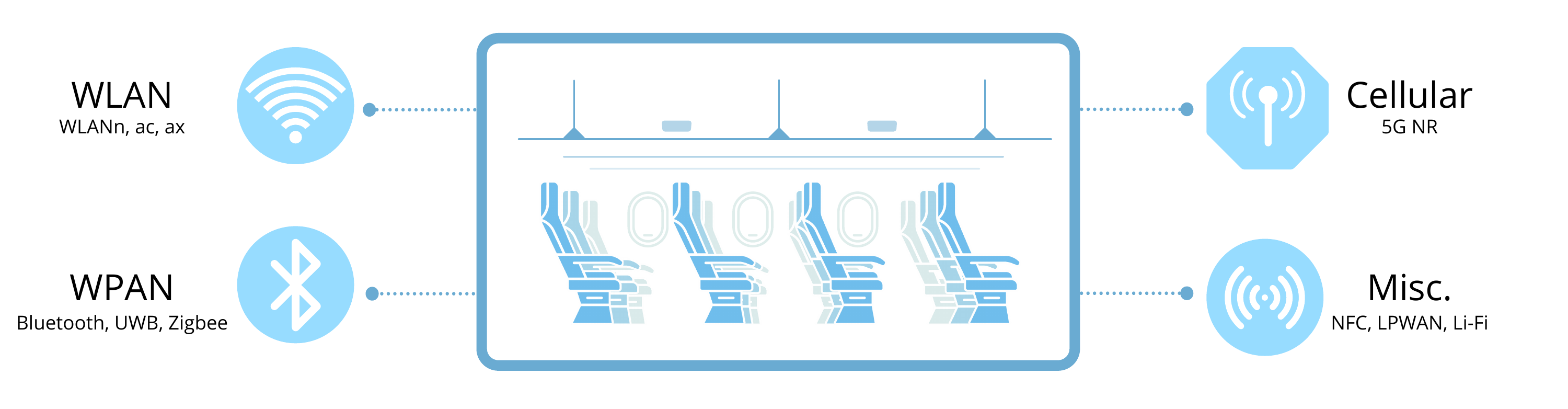}
    \caption{Overview of IPS technology candidates for location-based services within the connected cabin.}
    \label{fig:technologies}
\end{figure}

Possible technologies for radio-based positioning in the aircraft cabin are presented in Figure~\ref{fig:technologies}. These radio technologies differ by their center frequency, bandwidth, used waveform and other signal parameters that affects positioning performance. These technologies can be mainly categorized as cellular (LTE, 5G NR), Wireless Local Area Network (WLAN), Wireless Personal Area Network (WPAN), such as Bluetooth Low Energy (BLE), Zigbee, UWB and other technologies.

\subsection{Radio Channel and Modeling}


The performance of wireless communication and localization is heavily dependent on the radio channel and various propagation effects. The three basic propagation phenomena for the interaction of the radio wave with objects in the environment are reflection, scattering and diffraction. These phenomena lead to large-scale and small-scale channel characteristics of the radio propagation channel including path loss, shadowing and multipath fading. 

Existing channel models take these effects into account and could be categorized into deterministic, stochastic, and hybrid approaches. Stochastic models use statistical approximations in form of statistical distributions to characterize the received signal and channel characteristics such as path loss, delay spread, multipath components (MPCs) and the channel impulse response (CIR). These models are based on measurements and empirical observations in specific types of environments (rural, urban, indoor, micro, macro, etc.). The channel is modeled at a random location of the sensor in a defined type of environment, so there is no spatial consistency between different sensor locations. Due to the low complexity of stochastic models and the suitability to model a communication channel, there are used for example for the 3GPP TR 38.901 channel model \cite{3gpp2019technical}. 

Hybrid channel models often extend stochastic models by taking the geometry into account. For example, quasi-deterministic channel models compute the dominant MPC with a highly simplified environment map and add clusters of stochastically modeled MPCs to the model \cite{hanTerahertzWirelessChannels2022}. For geometry-based stochastic models the most widely used academic simulator is QuaDRiGa \cite{jaeckel2017quadriga}.

Deterministic models approximately solve the Maxwell's equations. The models achieve high accuracy, but require detailed knowledge of the geometry of the environment, the electromagnetic properties of the materials, and the spatial position of the sensors. The propagation paths or rays are calculated based on the channel characteristics and the position of reflectors in the environment. But such models are very computationally intensive. The signal parameters of every propagation path are usually computed using geometrical optics with two different methods: ray tracing and ray launching. \cite{hanTerahertzWirelessChannels2022}

Deterministic channel models consider the desired environment and sensor positions, thus there are spatial consistent. This spatial consistency is crucial to evaluate localization with indoor positioning systems (IPS) in a specific environment such as the aircraft cabin.

3D ray tracing computes all propagation paths or rays between a given transmitter and receiver position by tracing back from the receiver to the transmitter (cf. Figure~\ref{fig:Modell_Flugzeugkabine}). Various propagation phenomena and interactions, including reflection, diffraction, and scattering, are taken into account. By tracing the rays, the electrical lengths, amplitudes, and phases of the individual propagation paths is calculated. Thus, other channel characteristics, such as signal energy, delay spread, and the CIR can be determined. Ray launching on the other hand sends rays in all directions from the transmitter and looks which ones survive and arrive at the receiver. This procedure has a positive effect on the calculation time compared to ray tracing, especially in scenarios with one transmitter and many possible receiver positions, for example in the form of a grid. For this reason, we used ray launching as deterministic algorithm to compute all propagation paths and estimation of the signal parameters for our simulation toolchain presented in section~\ref{sec:methods}. 


\section{Methodology}
\label{sec:methods}

\subsection{Overview}



\begin{figure}[ht]
    \centering
    \includegraphics[width=\linewidth]{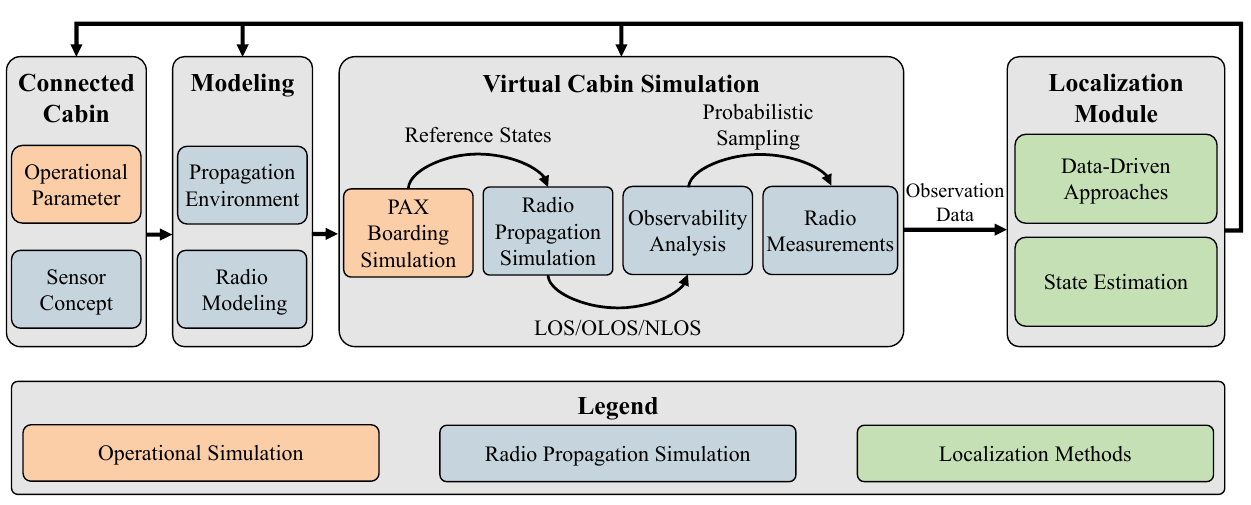}
    \caption{High-level flowchart of the proposed radio localization simulation approach.}
    \label{fig:flowchart}
\end{figure}

This work proposes a radio-positioning simulation approach, which aims at providing an universal and adaptable environment for investigating operational applications enhanced by location information within the connected cabin. As discussed before, for this application a simulation approach is required, which addresses the shortcomings of conventional investigations and simulation approaches concerning radio-based positioning. In this light, we present three main pillars, which constitute a framework for application-driven location-based services within the cabin. At first, operational requirements need to be addressed. This includes different use cases, the types of locatable objects and functional respectively quantitative requirements. The operation simulation allows the derivation of reference states from the operational inputs, e.g. a boarding sequence, where each PAX is associated to a reference position at each time step. Based on this, a scene-specific deterministic radio prapagation simulation can be performed, allowing to precisely derive channel information (e.g. reflections or LOS/NLOS) with respect to a specific environment and the operational inputs. This approach enables a spatial consistency of the radio simulation. However, only channel specific influences are considered. Lastly, a localization module needs to be implemented, which derives location information of the objects of interest and can represent both conventional state estimation or even data-driven approaches. Figure~\ref{fig:flowchart} summarizes a high-level overview on the embedding of the location simulation and how it can be incorporated within an operational scenario, while highlighting the pillars of the approach.

The main scope of the presented work is the adaptation of radio propagation simulation (as shown in ivory in Figure~\ref{fig:flowchart}). The developed approach utilizes both deterministic ray-tracing for spatial scene analysis, which is conventionally applied for network coverage analysis, and adaptive stochastic error sampling, partially based on empirical findings. In contrary to previously mentioned works, this spatial consistency is achieved by adopting the observability results from deterministic simulation, here Line-of-Sight, Obstructed Line-of-Sight or Non-Light-of Sight (Los, OLOS, NLOS) along additional parameters such as the channel impulse response or the delay spread. The obtained visibility constellations are then again used to further consider stochastic error influences with specific noise and error distributions for each propagation case as detailed in Section~\ref{ssec:ranging_simulation}. Based on these inputs, different localization methods can be trained and applied.

Since UWB-based positioning systems have previously been discussed for applications within the connected cabin, e.g. \cite{geyerPreciseOnboardAircraft2022}, we will also apply our simulation to the specifics and physical layer properties of UWB. In unison, we apply the principle of two-way-ranging as specified within the UWB standard \cite{Karapistoli2010OverviewUWBStandards}.

While not being the main focus of the work, we still present the incorporation of both operational simulation and localization methods (orange and green in Figure~\ref{fig:flowchart}) in order to emphasize the overall toolchain and interdependent principles of each pillar.

\subsection{Radio Propagation Simulation}

\begin{figure}[ht]
    \centering
    \includegraphics[width=.825\linewidth]{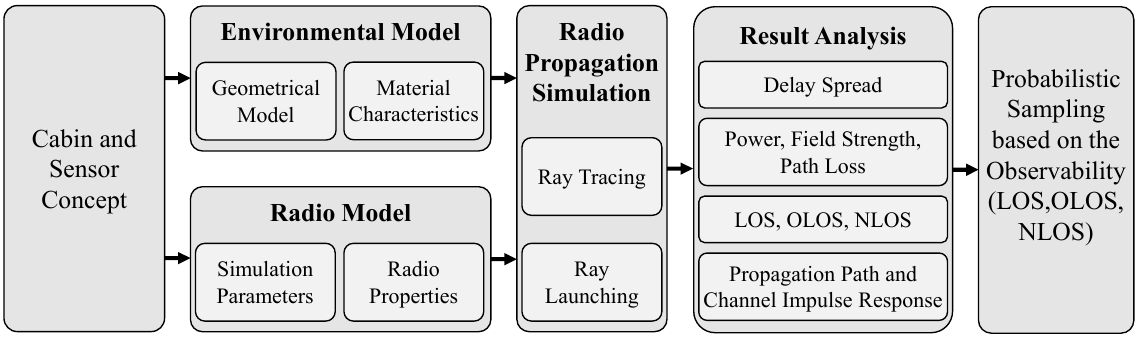}
    \caption{Processing for the radio propagation simulation with inputs, predictions models and results.}
    \label{fig:rps_flowgraph}
\end{figure}

For deterministic radio propagation simulation, various inputs and prerequisites to the modeling are necessary in order to subsequently perform the simulation based on the special propagation models and to derive the necessary channel parameters \cite{usslerDemoDeterministicRadio2022}. Figure~\ref{fig:rps_flowgraph} shows the processing for the deterministic radio propagation simulation to compute the visibility states (LOS/OLOS/NLOS) for the probabilistic sampling and the localization.

\begin{figure}[ht]
	\centering
        \begin{subfigure}[b]{1\textwidth}
		\centering
		\includegraphics[width=\textwidth]{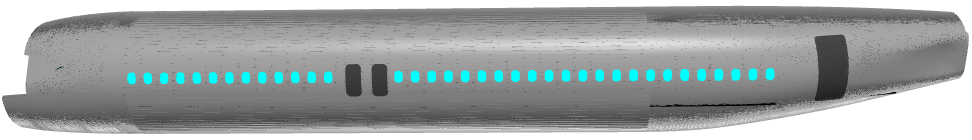} 
		\caption{}
		\label{fig:Modell_Flugzeugkabine_aussen}
	\end{subfigure}%
        \vspace{0.3cm}
	\begin{subfigure}[b]{0.47\textwidth}
		\centering
		\includegraphics[width=\textwidth]{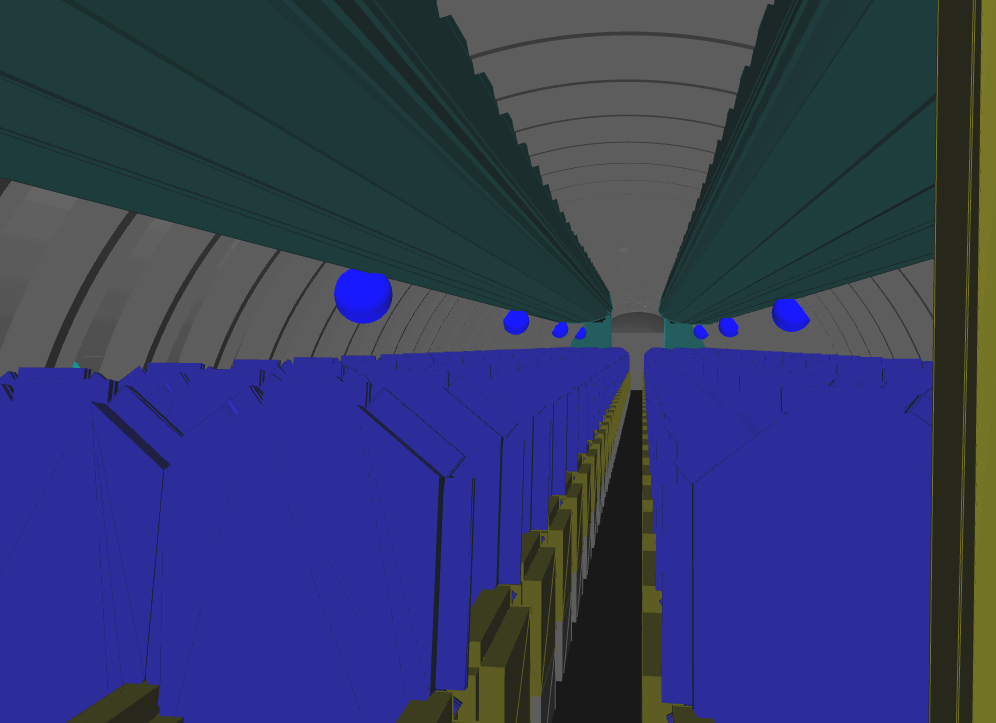}
		\caption{}
		\label{fig:Modell_Flugzeugkabine_innen}
	\end{subfigure}
        \hspace{0.2cm}
        \begin{subfigure}[b]{0.47\textwidth}
            \centering
            \includegraphics[width=\linewidth]{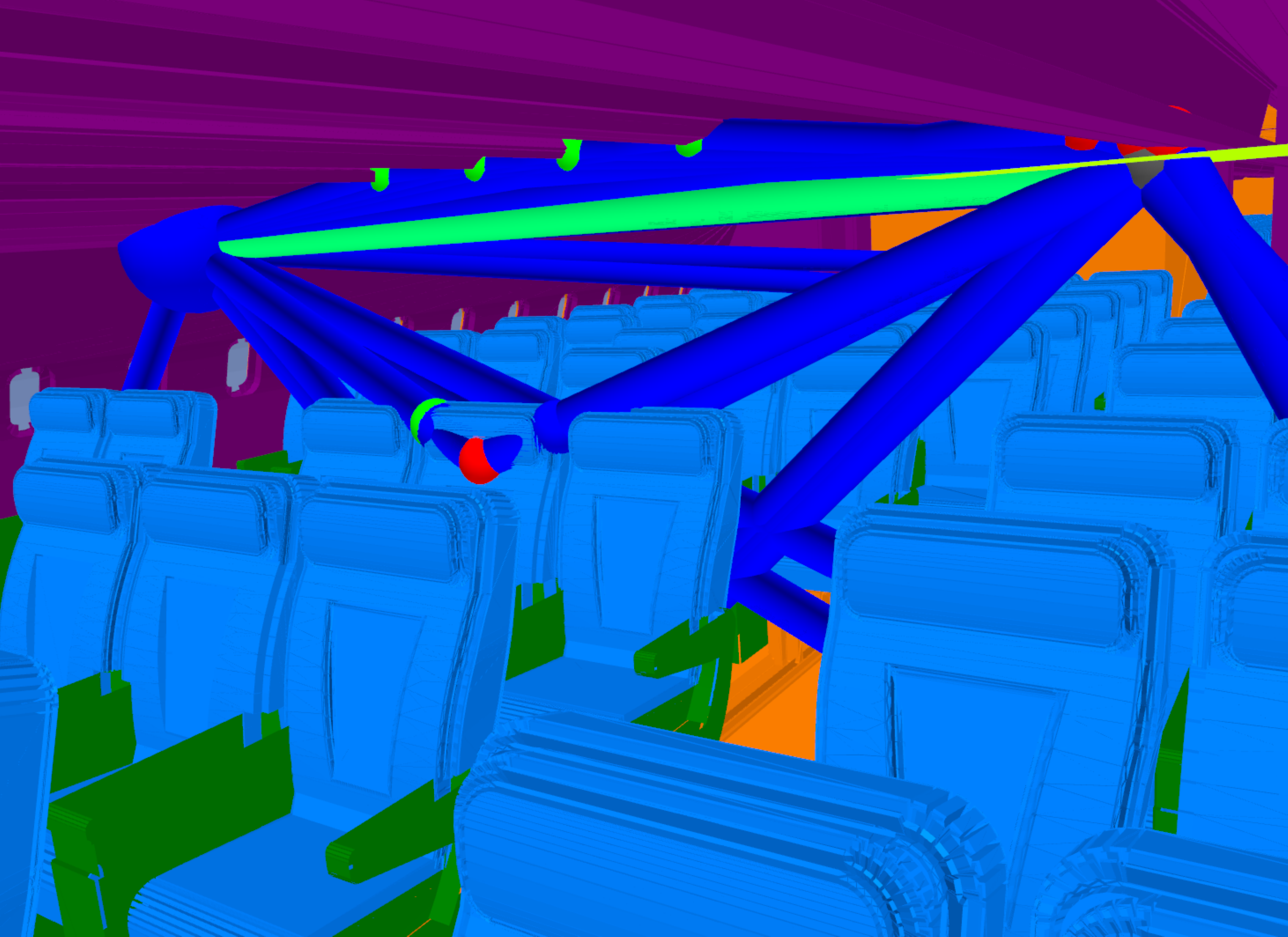}
            \caption{}
            \label{fig:simulation_ray_tracing_example}
        \end{subfigure}
	\caption{CAD-Model of the Airbus A320 aircraft cabin: \textbf{(a)} exterior, \textbf{(b)} interior, \textbf{(c)} exemplary result of 3D ray-tracing.}
	\label{fig:Modell_Flugzeugkabine}
\end{figure}

The inputs for the simulation include a detailed environment model that includes both the geometry of the objects and their electromagnetic properties. For this purpose we used a CAD model of the aircraft cabin with an associated database of material properties for different frequencies. The CAD model is an Airbus A320 including all seats, luggage compartments, but no actual PAX (cf. Figure~\ref{fig:Modell_Flugzeugkabine}). The electromagnetic properties include the relative permittivity, relative permeability and roughness of the various surfaces and objects in the model. From these properties it is possible to calculate the attenuation of the reflected wave and the loss due to transmission through the object with a defined thickness. \cite{usslerDemoDeterministicRadio2022}

The radio model consists of the signal parameters of the desired radio technology, the specification of the antenna and other simulation parameters that parameterize the computation method. The radio technology is specified by a channel, the associated center frequency and the transmit power. The modeling of the antenna is most accurately done by a 3D antenna pattern. From the antenna pattern, the antenna gain, the half-power beam width and the size the sidelobes can be obtained. Further parameters of the simulation include the maximum number of reflections, transmissions and diffraction per ray. Also, the specified sensitivity of the receiver determines the termination of the calculation of individual propagation paths.

Another specification for modeling the radio link is the location of the transmitter and receiver antennas. All propagation paths or rays are computed between the respective locations bases on the channel model and used prediction method. Transmit antennas are normally located at a fixed position, where as in our cases the position of the receiver antenna is varying within an equidistant grid. Thus, the visibility results and channel parameters are calculated for all receiver locations in the grid and are available for localization methods.

Ray launching provides different results and channel parameters, which are based on the calculation of the propagation paths with their respective length, amplitude and phase \cite{Yun2015RayTracing}. These signal and channel parameters are listed in Table~\ref{tab:Ergebnisse_Simulation} and are conventionally used to plan and optimize the performance of communication networks, however can also be applied for radio-based localization systems.

\begin{table}[ht]
        \centering
	\caption{Radio channel simulation results, which fully characterize the propagation channel and are required for the localization methods.}
	\label{tab:Ergebnisse_Simulation}
        \renewcommand{\arraystretch}{1.1}
	\begin{tabular}{p{3cm}p{10.5cm}}
    	\toprule
    	\textbf{Results}       & \textbf{Description}  \\
    	\midrule
    	Delay                 & Delay of the respective ray.                                                                                                    \\
    	Delay Spread                    & Delay spread of all propagation paths either as Excess Delay Spread or RMS Delay Spread.                                                                                                 \\
    	Path loss                     & Attenuation of the propagation path due to free-space path loss, reflections, scattering, diffraction, etc.                                                                                                        \\
    	Power / Field strength         & Calculated power and field strength at receiver based on the path loss of all propagation paths.                                                                                                                      \\
    	LOS/NLOS conditions & Differentiation whether there is a LOS, OLOS or NLOS condition between transmitter and receiver.                                                                           \\
    	Interactions with objects     &  Location of interaction and type of interaction (reflection, scattering or diffraction) of the propagation path with objects.    \\
    	Channel impulse response (CIR)                             & Delays and amplitudes of individual propagation paths plotted in the form of a power delay profile (PDP) with Dirac pulses. \\ \bottomrule	
    \end{tabular}
\end{table}

\subsection{Ranging Simulation}
\label{ssec:ranging_simulation}

\begin{figure}[htbp!]
    \centering
    \includegraphics[width=.65\linewidth]{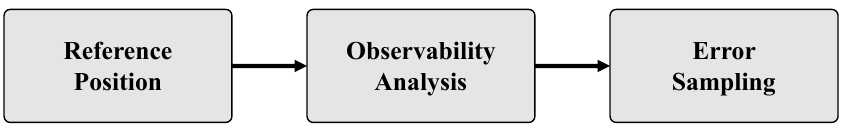}
    \caption{High-level error sampling procedure.}
    \label{fig:error_sampling}
\end{figure}

As previously mentioned, the summarized parameters in Table~\ref{tab:parameters_rps} can be further used for additional error sampling simulation. Specifically, stochastic sampling based on a semi-empirical simulation model has been extensively studied in previous works by examining the ranging capabilities of UWB in a harsh environment \cite{Schwarzbach_2021_statistical}. In accordance, a collaborative ranging simulation in a smart parking environment has been performed using this approach \cite{Jung2022AutoPositioning}. Therefore, the semi-empirical sampling approach is only briefly outlined below and will be further enhanced in the following sections by combining it with radio propagation simulation. Unlike aforementioned works, this paper enhances the semi-empirical approach with the previously introduced deterministic scene analysis and therefore allows more spatial consistency of the preserved observation data. The necessary steps for this task are briefly depicted in Figure~\ref{fig:error_sampling}.

The reference position are given from the operational scenario to be examined (cf. Figure~\ref{fig:flowchart}) and can either be time-variant or time-invariant. For each measurement step (epoch), two parameters can be derived based on the reference position and given infrastructure:

\begin{itemize}
    \item Reference distance $\boldsymbol{d}$, which is defined as the 3D distance between the reference position $\boldsymbol{\mathrm{x}}_{\text{ref}}$ and the position of available anchors $\boldsymbol{\mathrm{x}}_{a}$: $\boldsymbol{d} = \left\|  \boldsymbol{\mathrm{x}}_{a} - \boldsymbol{\mathrm{x}}_{\text{ref}} \right\|_2$.
    \item Observability analysis to each available infrastructure node is estimated from the given radio propagation simulation.
\end{itemize}

\begin{figure}[htbp!]
    \centering
    \begin{subfigure}[b]{0.28\textwidth}
    \centering
    \includegraphics[width=\textwidth]{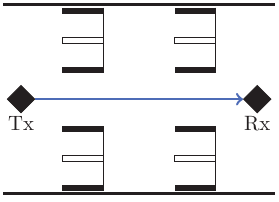}
    \caption{LOS}
    \label{fig:los}
    \end{subfigure}
    \hspace{0.5cm}
    \centering
    \begin{subfigure}[b]{0.28\textwidth}
    \centering
    \includegraphics[width=\textwidth]{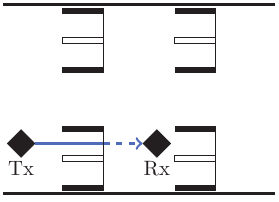}
    \caption{OLOS}
    \label{fig:olos}
    \end{subfigure}
    \hspace{0.5cm}
    \centering
    \begin{subfigure}[b]{0.28\textwidth}
    \centering
    \includegraphics[width=\textwidth]{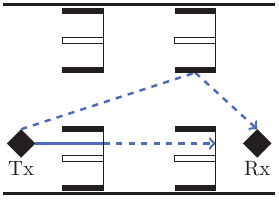}
    \caption{NLOS}
    \label{fig:nlos}
    \end{subfigure}
    \caption{Schematic examples of different radio propagation phenomena.}
    \label{fig:propagation_effects}
\end{figure}

For the latter there are generally four propagation phenomena observable \cite{qi2006analysis}, of which three are illustrated in Figure~\ref{fig:propagation_effects}:

\begin{itemize}
    \item \textbf{LOS}: The signal is received via the direct path between transmitter and receiver (cf. Figure~\ref{fig:los}).
    \item \textbf{OLOS}: The signal is still received via the geometrical direct path, however it has penetrated at least one structure on its way (cf. Figure~\ref{fig:olos}).  
    \item \textbf{NLOS}: The direct signal is blocked in this case, however the signal still received via a reflected path (cf. Figure~\ref{fig:nlos}).
    \item \textbf{Not receivable}: Given the propagation environment and the radio model, the signal attenuation can exceed the receiver sensitivity and therefore lead to a signal not being receivable at a given location. This effect is typically caused by high distances or many obstructing objects between transceivers.
\end{itemize}

In addition to these four propagation phenomena derived from radio propagation simulation, we additionally consider effects which are not induced by the propagation environment. At first, we consider general measurement noise modeled as additive white gaussian noise in order to quantify the overall precision of the measurement technology. These intrinsic factors are hardware-related and are empirically quantifiable \cite{Schwarzbach_2021_statistical}. Secondly, gross errors, also refered to as outliers, are substantial derivations between true and observable values. These effects can for example be caused by target confusion, network errors or electromagnetic interferences.

Given the aforementioned propagation scenarios and these additional error influences, we propose to model a ranging measurement ${r}$ as follows \cite{qi2006analysis, Schwarzbach_2021_statistical}:

\begin{align}
	r=\begin{cases} d+\p{101} & \text{if }p>\dfrac{d}{d_\mathrm{max}}\\ \emptyset & \text{else,}\end{cases}
	\label{equ:GenDaten1}
\end{align}

where the observation constitutes the true distance $d$ and the additive errors $\p{101}$. The error terms ${\p{101}}$ exhibit a linear dependence on distance, relative to an empirical maximum range denoted as $d_\text{max}$. The choice of this range is application and technology-specific. When the prescribed condition is not satisfied, we simulate a measurement failure denoted as $\emptyset$. Furthermore, the probability $p$ for occasional disturbances in measurements is represented as a Bernoulli experiment with $p\sim \mathcal{U}(0, 1)$. Empirical evidence from our previous study \cite{Schwarzbach_2021_statistical} has substantiated the presence of distance-dependent characteristics in errors. This phenomenon, in turn, exerts an influence on the success rate of simulated measurements. 

\begin{figure}[htbp!]
    \centering
    \scalebox{.9}{\input{fig/decision_tree}}
    \label{fig:decision_tree}
    \caption{Error sampling decision highlighting the underlying stochastic models.}
    \label{fig:sampling_scheme}
\end{figure}
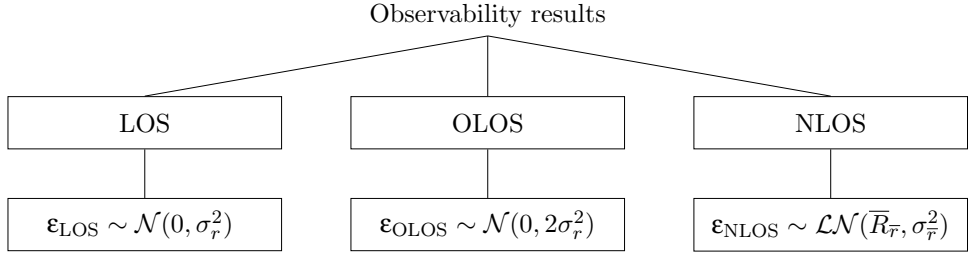

Depending on the observed propagation scenario, we sample our scenario-specific ranging error following the decision tree given in Figure~\ref{fig:sampling_scheme}, in which we introduce the following quantities:

\begin{itemize}
    \item $\p{101}_{\text{LOS}}$, $\p{101}_{\text{OLOS}}$ and $\p{101}_{\text{NLOS}}$ represent the ranging error for LOS, OLOS and NLOS scenarios respectively;
    \item $\mathcal{N}$ and $\mathcal{LN}$ denote a normal respectively a log-normal distribution;
    \item $\sigma_r^2$ represents the noise-related error term;
    \item $\overline{R}_r$ and $\sigma_{\overline{r}}^2$ represent the location and scale parameter of a $\mathcal{LN}$ distribution.
\end{itemize}

In addition, an empirical outlier probability $p_{\text{out}}$ is defined, which accounts for gross outliers. Outliers characterize the robustness of the measurement system and can be caused by a variety of influences, such as measurement confusions \cite{Schwarzbach_2021_statistical}. The quantity of this variable is again dependent on the environment and the measurement technology. Outliers are generally a major source for localization errors, especially in demanding scenarios. 

Therefore, a proper simulation of them is essential. Since outliers typically do not follow a generic stochastic model, we assume a uniform error distribution within physically measurable domain. Therefore outliers $\p{101}$ are additionally modeled as:

\begin{equation}
    \p{101}_{\text{OUT}}\sim\mathcal U(-d,d_\mathrm{max}-d).
\end{equation}

Further details on the stochastic sampling approach are given in \cite{Schwarzbach_2021_statistical}.

In summary, the following error types are considered within the simulation: \textbf{LOS} noise, \textbf{OLOS} noise, \textbf{NLOS} right-skewed noise, uniformly distributed \textbf{outliers} with respect to an outlier probability and \textbf{measurement failures} with respect to signal attenuation (maximum range) and random failures. Resulting overall error (residual) distribution derived from the presented simulation approach are presented in Section~\ref{sec:results}.

\subsection{Positioning}
\label{ssec:positioning}

In literature, a variety of applicable state estimation approaches for radio-based positioning exist. Most commonly, these can be divided in the following categories: Deterministic approaches use explicit observables and calculate geometric relationships between fixed infrastructure and mobile devices on this basis \cite{Stojmenovi_handbook_wsn_2005}. In addition, probabilistic approaches also use the aforementioned observables, but consider the system states as stochastic random variables and describe them as well as the observables by means of probability density functions \cite{Chen_Bayesian_Filtering_2003}. Lastly, the localization problem can also be solved by applying data-driven localization methods derived from machine learning and artificial intelligence \cite{Burghal2020ACS}.

Since the focus of the paper is on highlighting the simulation procedure in order to enable positioning applications (cf. Section~\ref{sec:methods}), this manuscript will not further elaborate applicable state estimation methods, but rather build up on existing frameworks allowing a discussion of the simulated effects on the localization process.

For dynamic systems, Recursive Bayes' Filter (RBF) pose a common framework for localization problems, including Kalman, Histogram or Particle Filters \cite{Chen_Bayesian_Filtering_2003}. In this work, we apply a pre-defined, equidistant grid representation of the state space, representing possible locations and allowing a multi-dimensional implementation of an Histogram Filter \cite{thrun2005probabilistic} or a point-mass filter  \cite{Bucy_digital_synthesis_non_linear_filters_1971}. This representation is applied for the given application because of two reasons. This non-paramtric filtering approach provides inherent robustness to non-gaussian error types compared to parametric filtering. With multiple locatable objects within the confined cabin space, the grid can be used as a uniform state space for all objects. When presented with various objects within the network, this approach reduces computational load compared to, for example, a Particle Filter.

Since this localization methods has been previously detailed for several radio-based localization problem, we only briefly outline the necessary calculation steps and refer to previous works for implementational details \cite{Schwarzbach2020TightIntegration, Jung2022AutoPositioning, Schwarzbach2023ION}.

For this method, also refered to as Grid Filtering, the domain of the state space representing realizations of the state vector $\boldsymbol{\mathcal{X}}$ is decomposed in a discrete and finite set of $M$-equidistant realizations $X_M$: $\text{dom}(\boldsymbol{\mathcal{X}})= \boldsymbol{X}_1 \cup \boldsymbol{X}_2 \cup \dots \boldsymbol{X}_M$.

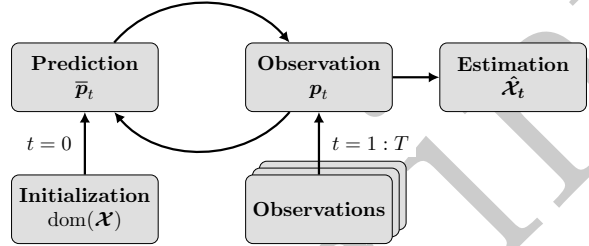
\begin{figure}[htbp!]
    \centering
    \scalebox{.7}{\input{fig/markov}}
    \caption{General RBF structure.}
    \label{fig:grid_flowchart}
\end{figure}

Given this state space initialization, the Grid Filter follows the general iterative RBF structure depicted in Figure~\ref{fig:grid_flowchart}, where the corresponding calculations are given in Equations~\ref{equ:notation1},  \ref{equ:notation2} and \ref{equ:notation3} \cite{thrun2005probabilistic}.

\begin{align}
    \overline{\boldsymbol{p}}_{t} &= \boldsymbol{p}_{t-1} \sum_m \boldsymbol{P}(\boldsymbol{\mathcal{X}}_{t, m}|{\boldsymbol{\mathcal{X}}}_{t-1}) \label{equ:notation1}\\
    \boldsymbol{p}_{t} &= \eta\; \overline{\boldsymbol{p}}_{t} \sum_m \boldsymbol{P} (\mathcal{\boldsymbol{\mathcal{Z}}}_{t} |\boldsymbol{\mathcal{X}}_{t, m}) \label{equ:notation2} \\
    \boldsymbol{\hat{\mathcal{X}}}_{t} &= \text{argmax}\;\boldsymbol{p}_{t},
    \label{equ:notation3}
\end{align}

where:

\begin{itemize}
    \item $\overline{\boldsymbol{p}}_{t}$ denotes the predicted belief following a motion model at timestep $t$,
    \item $\boldsymbol{p}_{t}$ denotes the resulting belief using the given observations modeled via 
    \item $\boldsymbol{P}(\mathcal{\boldsymbol{\mathcal{Z}}}_{t}|\boldsymbol{\mathcal{X}}_{t, m})$, which denotes the observation Likelihood with respect to the normalization constant~$\eta$
    \item $\boldsymbol{\hat{\mathcal{X}}}_{t}$ denotes the state estimation derived from maximizing the current belief.
\end{itemize}

\begin{figure}[htbp!]
    \centering
    \includegraphics[width=\linewidth]{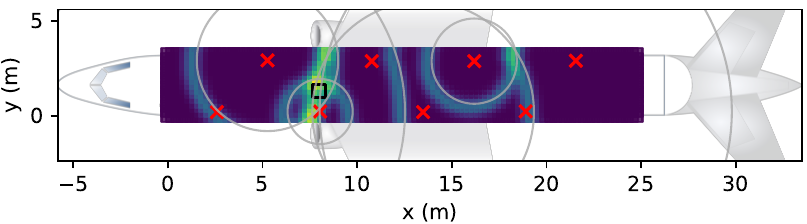}
    \caption{Exemplary output of the observation step of the Grid Filter using faulty ranging measurements. Anchors are depiced in red, reference point in black. The discrete state space is represented by an equi-distant grid (purple), where the likehoods of ranges (grey) are color-coded.}
    \label{fig:grid_state}
\end{figure}

Figure~\ref{fig:grid_state} depicts an exemplary output of the applied state estimation framework. Here, the given state space is shown as the - mostly - purple discrete grid. Reference points (red) perform faulty ranging measurements (grey) in accordance with Section~\ref{ssec:ranging_simulation}, which influence the grid likelihood as indicated by the yellow areas, which represent a higher likelihood. Finally, the reference position is indicated in black.

\section{Simulation and Results}
\label{sec:results}

\subsection{Simulation Overview}

In accordance with previous works, we constitute our simulation within an Airbus A320 model \cite{schultz_2018_implementation, Schwarzbach_2020_Covid19}. In total, three scenarios are covered as summarized in Table~\ref{tab:test_cases}. 

\begin{table}[htbp!]
\centering
\caption{Test cases on testbed for automated and connected driving.}
\begin{tabular}{@{}llll@{}}
\toprule
    & Name              & Description         & Samples                \\ \midrule
I   & Static, above seat & Antenna height 112cm & 5.000 positions, 40.000 ranges \\
II  & Static, below seat & Antenna height 70cm     & 5.000 positions, 40.000 ranges \\
III & Dynamic, 148 PAX  & Boarding process    & 40.184 positions, 321.472 ranges                           \\ \bottomrule
\end{tabular}
\label{tab:test_cases}
\end{table}

At first, two static scenarios are examined in order to achieve a reproducible scenario for different receiver heights. Scenario I refers to a receiver height slightly above the seats within the cabin, this generally leads to a high percentage of LOS propagation scenarios, while scenario II applies a receiver height indicated a position below the seat, e.g. for the localization of life vests \cite{Ninnemann_2022_multipath}. Both scenario I and II refer to an empty cabin without PAX. Scenario III then covers an entire boarding process, as previously described in \cite{schultz_fast_2018}. In total, 148 PAX are incorporated in the boarding sequence, which are continuously tracked during the process. 

Concerning the radio propagation simulation, ray launching is used to calculate radio parameters for all possible receiver locations within the cabin represented by an equi-distant grid. To limit the computational effort and time the maximum number of interactions with objects is set to two and the grid of possible receiver positions has a resolution of \SI{10}{\centi \meter}. The simulation is performed with the software \textit{Altair WinProp 2022.3} \cite{AltairWinProp2022}. Table~\ref{tab:parameters_rps} lists the parameter set used for the radio propagation simulation in the aircraft cabin. 

\begin{table}[htbp!]
\centering
\caption{Parameters and Inputs of the Radio Propagation Simulation with \textit{Altair WinPop 2022.3}.}
\label{tab:parameters_rps}
{\renewcommand{\arraystretch}{1.1}
    \begin{tabular}{p{3cm}p{5cm}p{5cm}}
    \toprule
                                           & Parameter                      & Value                                                                                                                                                                                                                     \\ \midrule
    Geometric Model                  & CAD Model                      & Airbus A320                                                                                                                                                                                                               \\ \midrule
    \multirow{4}{*}[-40pt]{\parbox{2cm}{Radio Properties}}                      & Center Frequency                      & \SI{6500}{\mega \hertz} (UWB Channel 5)                                                                                                                                                                                                  \\
                                           & TX Power                       & \SI{-14,7}{\dBm}                                                                                                                                                                                                                 \\
                                            & Antenna                       & Omidirectional Antenna                                                                                         \\
                                           & Position of Anchors [\si{\meter}]           & \begin{tabular}[c]{@{}l@{}}A1: -1.36;11.11;1.50\\ A2: 1.37;13.77;1.50\\ A3: -1.33;16.55;1.50\\ A4: 1.33;19.28;1.50\\ A5: -1.36;21.99;1.50\\ A6: 1.33;24.69;1.50\\ A7: -1.32;27.41;1.50\\ A8: 1.35;30.06;1.50\end{tabular} \\ \midrule
    \multirow{5}{*}{\parbox{2cm}{Simulation Parameters}} & RX Grid Resolution & \SI{10}{\centi \meter}                                                                                                                                                                                                                     \\ 
                                           & RX Prediction Plane Heights    & \SI{70}{\centi \meter} and \SI{112}{\centi \meter} above floor                                                                                                                                                                                                       \\ 
                                           & Prediction Model                          & Ray Launching                                                                                                                                                                                                             \\
                                           & Max. \# of Reflections     & 2                                                                                                                                                                                                                         \\
                                           & Max. \# of Transmissions    & 2       \\             & Max. \# of Diffractions    & 0       \\                                                                                                                                                                                         
    \bottomrule
    \end{tabular}
    }
\end{table}

\subsection{Visibility Analysis}

\begin{figure}[htbp!]
    \centering
    \includegraphics[width=\linewidth]{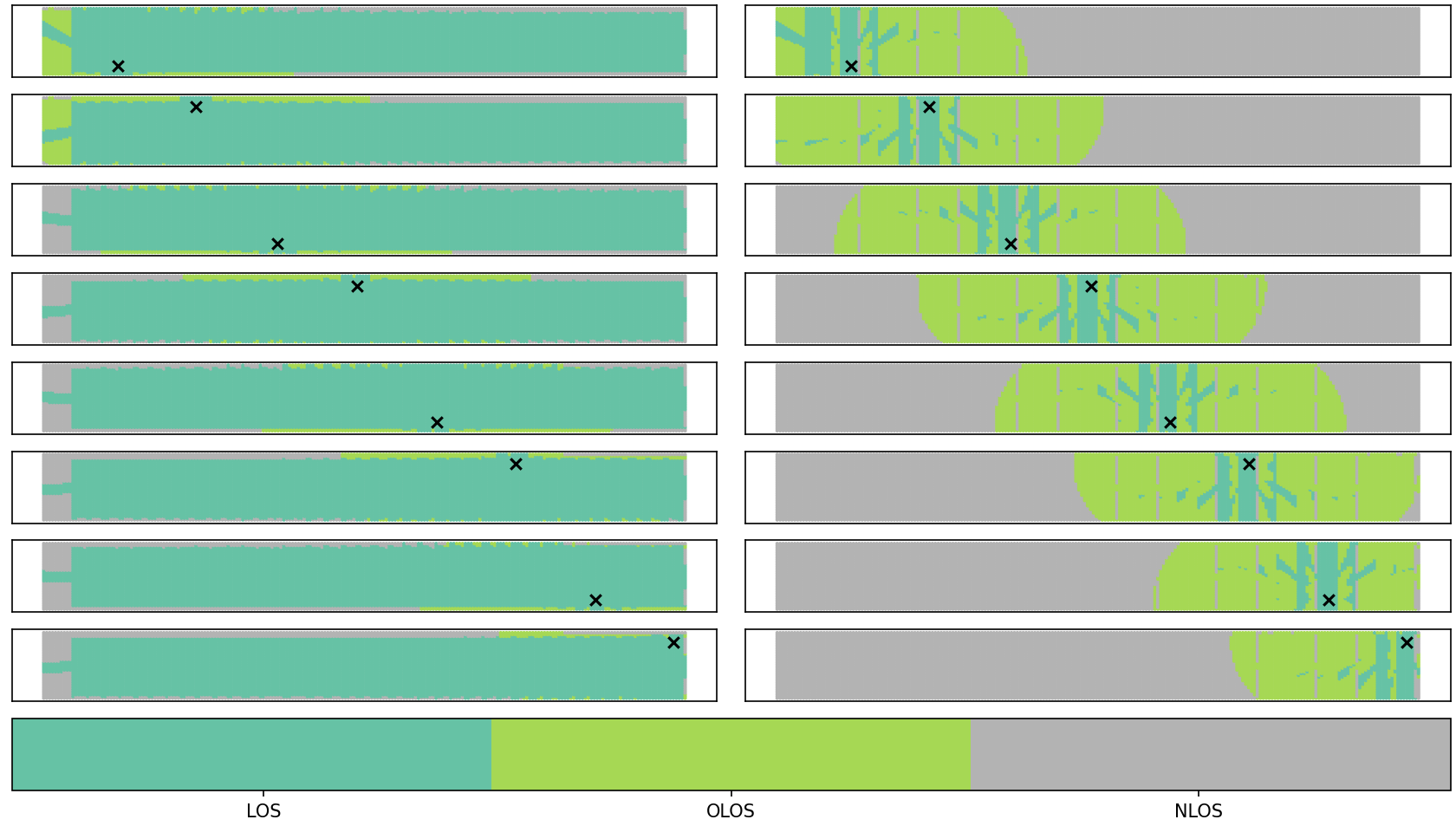}
    \caption{Visibility Analysis provided by radio propagation simulation: Scenario I (left) and scenario II (right) for the entire reception plane. The visibility states are color-coded with respect to the given colorbar.}
    \label{fig:los_states}
\end{figure}

The first result received from the deterministic radio simulation is the observability between anchors and the reference positions. In order to illustrate these constellations, the visibility states for scenario I and II are compared in Figure~\ref{fig:los_states}. For each anchor indicated in black, the observation state throughout the grid is given by the respective color. 

Given the results in Figure~\ref{fig:los_states}, the difference in relative antenna height between anchors and the reception plane is apparent. While scenario I above the seats comprises LOS reception in an empty cabin, scenario II constitutes hardly any LOS states due to the obstruction of objects between anchors and the reference plane. For scenario II, we also considered a distant-dependent threshold between OLOS and NLOS observations, as additional signal attenuation due to longer travel distances for the OLOS path would lead to a higher likelihood for NLOs reception. In the given case, this threshold was empirically set to $\SI{6}{\meter}$.

\subsection{Ranging simulation}

\begin{figure}[htbp!]
		\centering
        \includegraphics[width=.9\linewidth]{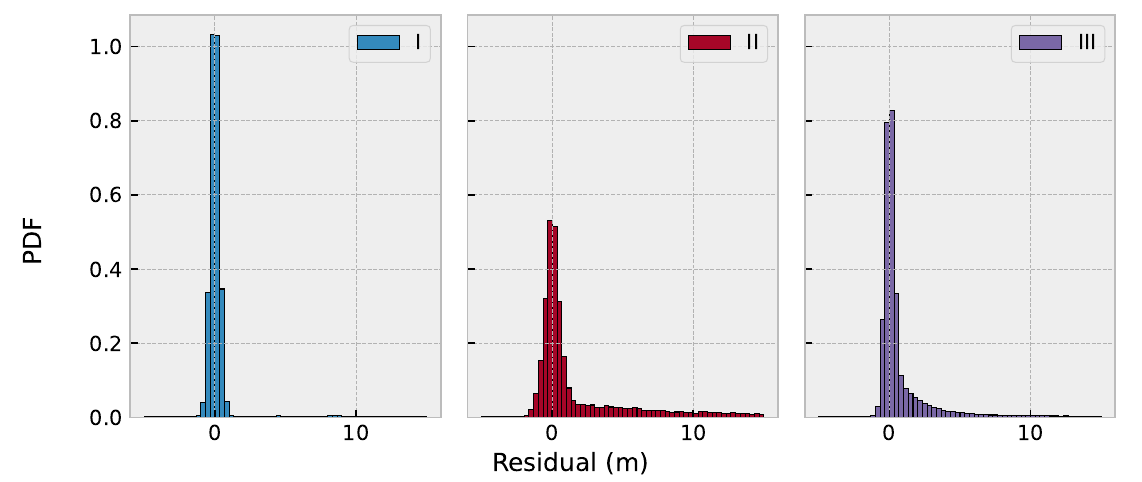}
        \caption{Residuals for simulated scenarios. Values are clipped at $[-5, 15]$ m.}
        \label{fig:measurement_residuals}
\end{figure}

 Based on the derived results from the visibility analysis, the stochastic error sampling as detailed in Section~\ref{ssec:ranging_simulation} is performed. In order to provide insight on the different visibility states, Figure~\ref{fig:measurement_residuals} depicts the ranging residuals (measurement errors) for each individual scenario. In order to highlight the underlying data, the corresponding quantities are briefly outlined in Table~\ref{tab:number_measurements}, which further highlights the presence of both measurement failures in the confined cabin area ($\SI{21.4}{\percent}$ to $\SI{22.5}{\percent}$ across all three scenarios) as well as the presence of measurements outliers, which are not explicitly shown in Figure~\ref{fig:measurement_residuals} due to readability. \vspace{0.5cm}

\begin{table}[htbp!]
\caption{Qunatities of ranging measurements for each scenario}
\begin{tabular}{@{}cccc@{}}
\toprule
Szenario & Total Measurements & Valid Measurements & Depicted Measurements \\ \midrule
I        & $40.000 \; (\SI{100}{\percent})$             & $31.437 \; (\SI{78.6}{\percent})$             & $30.269 \; (\SI{75.7}{\percent})$              \\
II       & $40.000 \; (\SI{100}{\percent})$             & $31.469 \; (\SI{78.7}{\percent})$             & $28.353 \; (\SI{70.9}{\percent})$                \\
III      & $322.376 \; (\SI{100}{\percent})$            & $233.569 \; (\SI{72.5}{\percent})$           & $224.683 \; (\SI{70.0}{\percent})$              \\ \bottomrule
\end{tabular}
\label{tab:number_measurements}
\end{table}


It is observable, that the measurement residuals follow distinct distributions influenced by the simulated scenarios. Scenario I constitutes a majority of LOS reception (as indicated by Figure~\ref{fig:los_states}), while still being prone to environmentally unrelated outliers and measurement failures. In contrast, scenario II is prone to a majority of comparably higher noise due to OLOS as well as manifold NLOS receptions. Lastly, caused by the presence of additional obstacles within the propagation environment, scenario III reveals a mix of the two previous scenarios. 

\subsection{Positioning results}

\begin{figure}[htbp!]
    \centering
    \includegraphics[width=\linewidth]{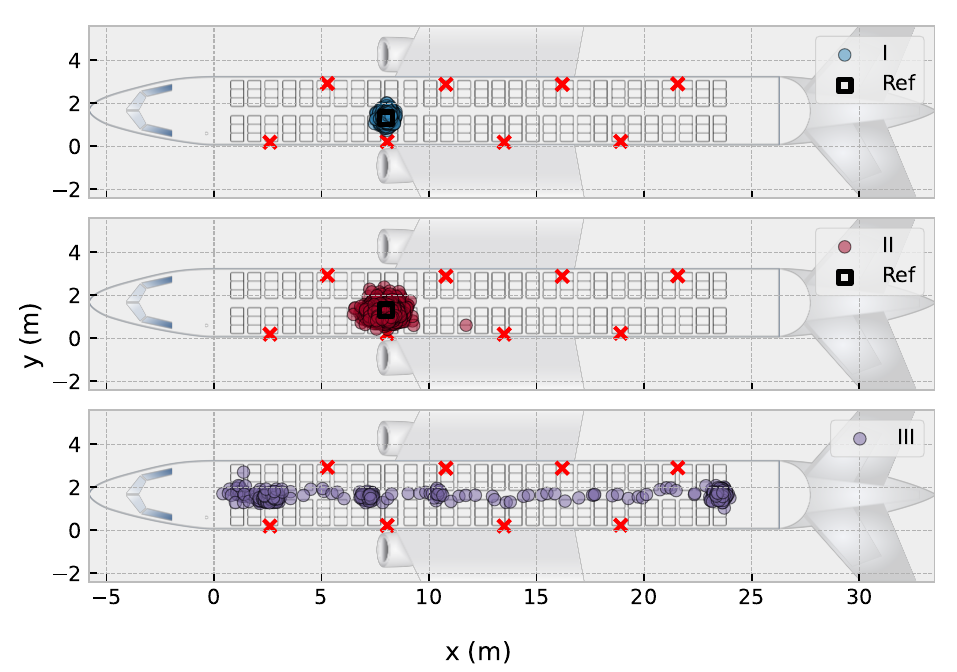}
    \caption{Qualitative positioning results for each simulation scenario: Scenario I and II represent the static receiver position at different heights, while for scenario III the trajectory of one exemplary PAX is depicted.}
    \label{fig:positioning_static}
\end{figure}

\begin{figure}[htbp!]
	\centering
        \begin{subfigure}[b]{0.49\textwidth}
		\centering
            \includegraphics[width=\linewidth]{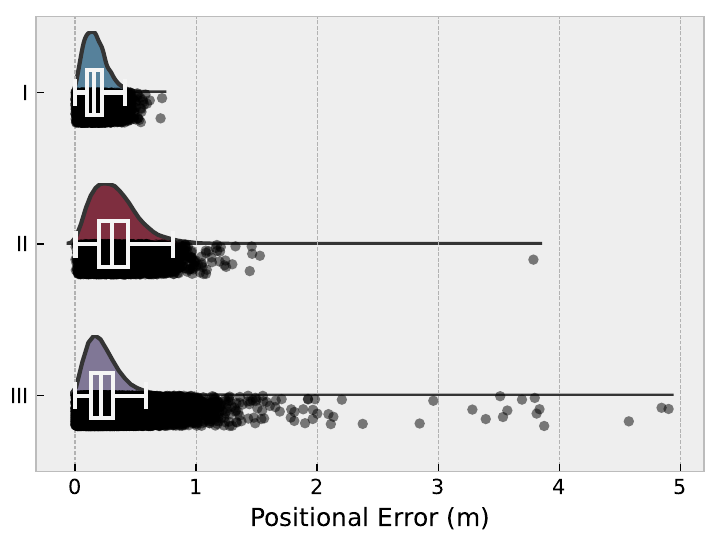}
            \caption{}
            \label{fig:viollinplots}
	\end{subfigure}
	\begin{subfigure}[b]{0.49\textwidth}
		\centering
		\includegraphics[width=\textwidth]{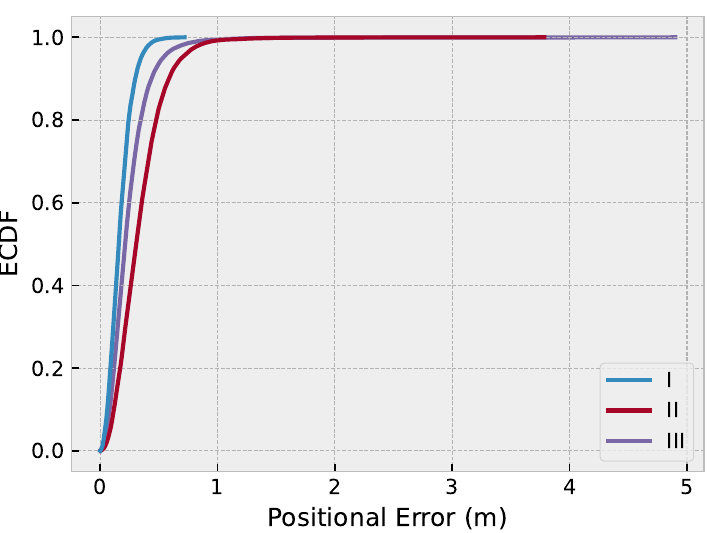}
		\caption{}
		\label{fig:cdf}
	\end{subfigure}
	\caption{Statistical evaluation of the positioning error for all three scenarios \textbf{a)} Raincloud plot and individual error points. \textbf{b)} empirical cumulative density function (ECDF). Positional errors are clipped at $\SI{5}{\meter}$.}
    \label{fig:positioning_results}
\end{figure}

At last, the discussion of derivable positioning results based on the introduced positioning method from Section~\ref{ssec:positioning} is presented. Figure~\ref{fig:positioning_static} depicts the qualitative localization results for the simulated scenarios. Furthermore, a quantitative analysis of the positioning results is conducted. For this the $\lvert \lvert \cdot \rvert \rvert_2$ norm between the estimated and reference positions is utilized as error metric $\mathcal{Q}$ \cite{iso_information_technology_2016}. For this matter, Table~\ref{tab:statistics} summarizes relevant statistical measures for the positioning accuracy, while Figure~\ref{fig:positioning_results} depicts these measures as well as an empirical cumulative density function of the positioning error. The comparison of scenario I and II in Figure~\ref{fig:positioning_static} clearly reveals the influence of different measurement error distributions on the positioning results, as the right-skewed error distribution in scenario II also leads to a higher scattering of positioning results. This is also supported by Figure~\ref{fig:positioning_results}, where scenario I yields an average positioning error of $\overline{\mathcal{Q}}_{\text{I}} = \SI{0.17}{\meter}$ compared to $\overline{\mathcal{Q}}_{\text{II}} = \SI{0.34}{\meter}$ for scenario II. The overall descriptive statistics for all scenarios are summarized in Table~\ref{tab:statistics}, including average $\bar{\mathcal{Q}}$ (root mean square error), median $\tilde{\mathcal{Q}}$ and the error variances $\sigma_{\mathcal{Q}}^2$. In addition, the $\sigma$ quantiles and the $25, 50, 75$ percentiles of the error distributions are given.

\begin{table}[htbp!]
\renewcommand{\arraystretch}{1.}
\centering
\caption{Quantitative evaluation scenario I, II and III.}
\begin{tabular}{@{}cccccccccc@{}}
\toprule
\multirow{2}{*}{Measure} & $\bar{\mathcal{Q}}$    & $\tilde{\mathcal{Q}}$  & $\sigma_{\mathcal{Q}}^2$ & \multicolumn{3}{c}{Quantile $[ \SI{}{\meter} ]$          }                         & \multicolumn{3}{c}{Percentile $[ \SI{}{\meter} ]$} \\ 
                         & $[ \SI{}{\meter} ]$ &  $[ \SI{}{\meter} ]$ & $[ \SI{}{\meter} ]^2$               & $\sigma$ & $2\sigma$ & $3\sigma$ & $25$          & $50$          & $75$         \\ \midrule
I                   &   $0.17$      &   $0.16$      &    $0.01$                                      &   $0.20$                    &     $0.35$                   &      $0.45$                  &   $0.10$          &     $0.16$        &    $0.23$       \\
II                  &   $0.34$      &  $0.31$       &       $0.04$                                   &      $0.40$                 &     $0.69$                   &         $0.93$               &  $0.19$           &    $0.31$         &     $0.44$       \\ 
III                  &   $0.29$      &  $0.21$       &       $0.69$                                   &      $0.28$                 &     $0.54$                   &         $0.89$               &  $0.13$           &    $0.21$         &     $0.31$       \\

\bottomrule
\end{tabular}
\label{tab:statistics}
\end{table}

Furthermore, scenario III is briefly discussed. Figure~\ref{fig:positioning_static} reveals clustering points in different areas, where the PAX has to wait before moving on due to other PAX. These occurrences can be monitored for operational planning and turnaround estimation respectively the optimization of the overall boarding strategy.

While Figure~\ref{fig:positioning_static} only depicts the estimated trajectory for one PAX, the given statistical measures include the positional errors for all 148 simulated PAX during the boarding sequence. From the available results, it can be inferred that the presence of PAX within the cabin has an additional negative impact on visibility, and consequently, on ranging accuracy.

\section{Conclusion \& Research Directions}

Radio propagation simulation is an ubiquitous tool in various engineering fields, predominantly used for coverage and network analysis. However, it also provides numerous advantages for radio-based location applications. However, single-handedly applied deterministic, empirical and stochastic models each have dedicated disadvantages for simulating the necessary inputs for radio-based localization. Therefore, this paper proposes the integrated combination of the aforementioned approaches in order to provide spatially consistent yet enabling alternating residual distributions. Due to the adaptability of both the deterministic and the stochastic simulation components, different scenarios and error sampling distributions can easily be applied and assessed. The framework also enables a cheap setup optimization, for example geometrical layout of anchors, and parameter tuning, for example for outlier detection.

The implemented simulation toolchain can be used not only to evaluate the performance of localization algorithms based on visibility states but is also scalable in terms of evaluation opportunities. These opportunities encompass assessing other signal parameters, such as RSSI, Delay, or CIR (as shown in Table~\ref{tab:Ergebnisse_Simulation}), and evaluating different propagation environments and radio technologies, including specific antenna hardware.

In summary, the presented results encompass three scenarios, each yielding different visibility states, resulting in demanding and partially right-skewed error distributions based on the proposed sampling distributions. This further underscores the need for robust positioning and effective outlier mitigation strategies.

We want to further emphasize, that the paper only covers one specific network constellation, one radio technology and a specific distance measurement principle. Derived from that, a fixed set of proposal distributions and one localization scheme is presented. However, due to the nature of the toolchain, manifold of the expects can be altered and compared. This also allows the simulation toolchain to be applied for the implementation and evaluation of data-driven localization approaches and machine leaning in general by addressing the data generation bottleneck. Based on the channel model the signal parameters used for ranging and simulation could be obtained by the deterministic radio propagation simulation with ray tracing or ray launching for a desired environment and use case \cite{michlerPotentialsDeterministicRadio2023}. This also aligns with research directions for the newest generation of radio positioning systems \cite{Schwarzbach2022Enabling}.





\section*{Funding}

\begin{tabular}{p{10cm} p{5cm}}
This work has been funded by the German Federal Ministry for Economic Affairs and Climate Action (BMWK) following a resolution of the German Federal Parliament within the projects CANARIA (FKZ: 20D1931C) and INTACT (FKZ: 20D2128D). & 
\hspace{-0.5 cm} \raisebox{-0.8\height}{\includegraphics[width=0.6\linewidth]{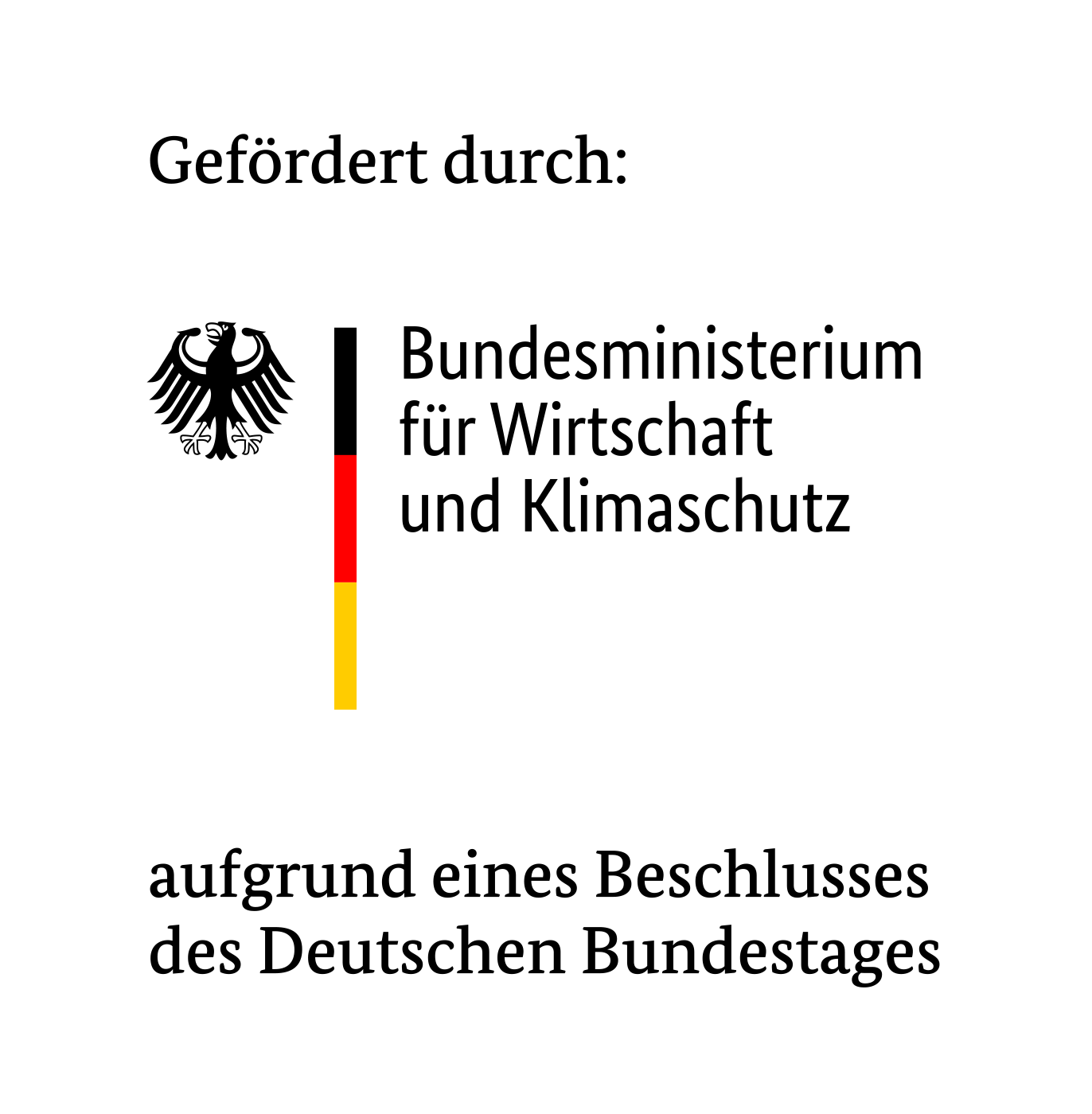}} \\
\end{tabular}





\section*{Dataset}

The dataset consist of three .csv files containing the visibility analysis from ray launching, the anchor states and the results of the probabilistic range sampling. The file \verb|visibility.csv| contain visibility states inside a grid with resolution of \SI{10}{\centi \meter} of the two simulated heights in the cabin. The columns of the file include the following data:
\begin{itemize}
    \item \verb|x|: position of the predicted grid cell in x.
    \item \verb|y|: position of the predicted grid cell in y.
    \item \verb|height|: height of the simulated prediction plane. Is used to investigate the viability of the tag for two different height in the cabin.
    \item \verb|anchor_id|: ID of the anchor for which the visibility is specified.
    \item \verb|visibility|: computed visibility state (\verb|LOS=1|/\verb|OLOS=2|/\verb|NLOS=3|) based on the radio propagation simulation with ray launching.
\end{itemize}

The second .csv file includes the sampled ranges based on the observability and is named \verb|ranges.csv|. Each line of the file contains the following data:
\begin{itemize}
    \item \verb|ref_pos_x|: static reference position of the PAX in x.
    \item \verb|ref_pos_y|: static reference position of the PAX in y.
    \item \verb|height|: height of the used prediction plane.
    \item \verb|anchor_id|: ID of the anchor for which the range is sampled.
    \item \verb|range|: sampled range between the anchor and the tag at the PAX. This range is used in the localization module to estimate the position of the PAX based on the simulated radio channel characteristics and visibility state from the radio propagation simulation.
\end{itemize}

Additionally the 3D position of the anchors in the cabin are provided in separate file called \verb|anchors.txt|. The dataset is open-access available to download on the research data repository OpARA of Dresden University of Technology: \textit{The submitted data is still under review at the research data management platform. The data has temporarily been stored at:} \url{https://github.com/PaulSchwarzbach/ewgt_2023.git}.


\bibliographystyle{tfcad}
\bibliography{lit}

\end{document}

%% file: fig/decision_tree.tex
\begin{tikzpicture}[
    every tree node/.style={align=center,anchor=north},
    level distance=1.5cm,
    level 1/.style={sibling distance=1.cm},
    level 2/.style={sibling distance=.5cm}
]
\tikzset{
  box/.style={
    draw,
    rectangle,
    minimum width=4cm,
    minimum height=2em,
    inner sep=1ex,
  }
}
\Tree [.{Observability results}
        [.\node[box]{LOS}; 
            [.\node[box]{$\p{101}_{\text{LOS}}\sim \mathcal{N}(0,\sigma_r^2)$  }; ]
        ]
        [.\node[box]{OLOS};
            [.\node[box]{$\p{101}_{\text{OLOS}}\sim \mathcal{N}(0,2\sigma_r^2)$}; ]
        ]
        [.\node[box]{NLOS};
            [.\node[box]{$\p{101}_{\text{NLOS}}\sim\mathcal{LN}(\overline{R}_{\overline{r}},\sigma_{\overline{r}}^2)$}; ]
        ]
    ]
\end{tikzpicture}

%% file: fig/markov.tex
\tikzset{%
  >={Latex[width=2mm,length=2mm]},
            base/.style = {rectangle, rounded corners, draw=black,
                           minimum width=2.5cm, minimum height=1.3cm,
                           text width = 2.5cm, fill=gray!25,
                           text centered},
  			standard/.style = {rectangle, rounded corners, draw=black,
                           minimum width=2.5cm, minimum height=1.3cm,
                           text width = 2.5cm, fill=gray!25,
                           text centered},
            line/.style={draw, very thick, ->},
            standard1/.style = {rectangle, rounded corners, draw=black,
                           minimum width=2.5cm, minimum height=1.3cm,
                           text width = 2.5cm, fill=gray!25,
                           text centered},
}
   
\begin{tikzpicture}[node distance=1.9cm,
    every node/.style={fill=white}, align=center]
  \node (start)             [standard]              {{\textbf{Initialization} \\ $\text{dom}(\boldsymbol{\mathcal{X}})$}};
  \node (predict) [base, above of=start, yshift=.6cm] {{\textbf{Prediction} \\ $\overline{\boldsymbol{p}}_{t}$}};
  \node (observe) [base, right of=predict, xshift=2.5cm] {{\textbf{Observation}\\$\boldsymbol{p}_{t}$}};
  
  \node (data) [standard] at (observe |- start) {{\textbf{Observations}}};
  
  \begin{pgfonlayer}{bg} 
   \begin{scope}
        \node (data2) [standard, xshift=0.26cm, yshift=.26cm] at (observe |- start) {};
        \node (data1) [standard, xshift=0.13cm, yshift=0.13cm] at (observe |- start) {};
       \end{scope}
 \end{pgfonlayer}

  \node (estimate) [standard1, right of=observe, xshift=1.75cm] {{\textbf{Estimation}\\$\boldsymbol{\hat{\mathcal{X}}_t}$}};

   \path[line] (start) -- node[midway,left, xshift=-0.1cm] {{$t=0$}} (predict);
   \path[line] (data) -- node[midway,right, xshift=0.1cm] {{$t=1:T$}} (observe);
   \path[line] (observe) -- (estimate);
    \path[line, <-] (predict) to[bend right=50](observe);
    \path[line] (predict) to[bend left=50](observe);
  \end{tikzpicture}